\documentclass{emulateapj}

\shorttitle{The {\it barless} $M_{\rm bh}$-$\sigma$ relation}
\shortauthors{Alister W.\ Graham}

\begin{document}

\title{Fundamental Planes, and the {\it barless} $M_{\rm bh}$-$\sigma$ relation, 
for Supermassive Black Holes}

\author{Alister W.\ Graham\altaffilmark{1}}
\affil{Centre for Astrophysics and Supercomputing, Swinburne University
of Technology, Hawthorn, Victoria 3122, Australia.}
\altaffiltext{1}{Corresponding Author: AGraham@astro.swin.edu.au}

\begin{abstract}  

The residuals about the standard $M_{\rm bh}$-$\sigma$ relation correlate with
the effective radius, absolute magnitude, and S\'ersic index of the host
bulge.  Although, it is noted here that the elliptical galaxies do not partake
in such correlations. Moreover, it is revealed that barred galaxies (with
their relatively small, faint, and low stellar concentration bulges) can
deviate from the $M_{\rm bh}$-$\sigma$ relation by $\delta \log M_{\rm bh}
\approx -0.5$ to $-1.0$ dex (their $\sigma$ values are too large) and generate
much of the aforementioned correlations.  Removal of the seven barred galaxies
from the Tremaine et al.\ set of 31 galaxies gives a ``barless $M_{\rm
bh}$-$\sigma$'' relation with an intrinsic scatter of 0.17 dex (cf.\ 0.27 dex
for the 31 galaxies) and a total scatter of 0.25 dex (cf.\ 0.34 dex for the 31
galaxies).  The introduction of a third parameter does not reduce the scatter.
Furthermore, removal of the barred galaxies, or all the disk galaxies, from an
expanded and updated set of 40 galaxies with direct black hole mass
measurements gives a consistent result, such that $\log(M_{\rm bh}/M_{\sun}) =
(8.25\pm0.05) + (3.68\pm0.25)\log [\sigma/200\, {\rm km\, s}^{-1}]$.  
The (barless) $\sigma$-$L$ relation for galaxies with
black hole mass measurements is found to be consistent with that from the SDSS
sample of early-type galaxies.
In addition the barless $M_{\rm bh}$-$\sigma$ relation, the $M_{\rm bh}$-$n$ 
relation, and the $M_{\rm bh}$-$L$ relation are shown to yield SMBH masses 
less than 2-4$\times 10^9 M_{\sun}$. 
%
%
%

\end{abstract}

\keywords{ black hole physics --- galaxies: bulges --- galaxies: fundamental
parameters --- galaxies: structure }

\section{Introduction}

Tight correlations between supermassive black hole (SMBH) masses and large
scale properties of the host bulges are interesting for two obvious reasons.
They enable us to predict SMBH masses in thousands of galaxies where the black
hole's sphere-of-influence is highly unresolved, and they provide clues to the
physical processes responsible for the co-evolution of black hole and host
bulge.  Recent endeavors have advocated relations involving not one but two bulge
parameters and therein claims of ``fundamental planes'', akin to the
Fundamental Plane for elliptical galaxies (Djorgovski \& Davis 1987;
Dressler et al.\ 1987).
The existence of such SMBH fundamental planes imply that current theories for the $M_{\rm
bh}$-$\sigma$ relation (Ferrarese \& Merritt 2000; Gebhardt et al.\ 2000) or 
the $M_{\rm
bh}$-$L$ relation (McLure \& Dunlop 2002; Marconi \& Hunt 2003; 
updated in Graham 2007), which do not include a third parameter, are
incomplete. 

This
article investigates the fundamental planes for SMBHs\footnote{The 
``fundamental plane of black hole activity'' involving radio core
luminosity and X-ray luminosity (Merloni et al.\ 2003; Falcke et al.\ 2004) is
not addressed here.}
involving the parameters 
$M_{\rm bh}$, $R_{\rm e}$, and $\langle \mu \rangle_{\rm e}$ (Barway \&
Kembhavi 2007), 
$M_{\rm bh}$, $R_{\rm e}$, and $\sigma$ (Marconi \&
Hunt 2003; de Francesco et al.\ 2006; Aller \& Richstone 2007; Hopkins et al.\ 
2007), and, for the first time, 
$M_{\rm bh}$, $\sigma$, and $n$.  
In Section~2 it is explained why a previous claim for a small `total' scatter
(0.19 dex) about the $M_{\rm bh}$-$R_{\rm e}$-$\langle \mu \rangle _{\rm e}$
plane was the result of a miscalculation.
In Section~3 it is revealed that the galaxies which deviate from the 
$M_{\rm bh}$-$\sigma$ relation, giving rise to the 
$M_{\rm bh}$-$\sigma$-$R_{\rm e}$, 
$M_{\rm bh}$-$\sigma$-$L$,  and 
$M_{\rm bh}$-$\sigma$-$n$ relations with less scatter than the 
$M_{\rm bh}$-$\sigma$ relation, are predominantly barred galaxies.  A ``barless 
$M_{\rm bh}$-$\sigma$'' relation, and an elliptical-only $M_{\rm
  bh}$-$\sigma$ relation, is subsequently constructed in Section~4 and found to 
heavily nullify the evidence for fundamental planes for SMBHs and their 
host bulges.  

Given the recent discussion in the literature about biases in 
the $M_{\rm bh}$-$\sigma$ and/or $M_{\rm bh}$-$L$ relation, and also in
the local sample of galaxies with direct SMBH mass measurements, these 
concerns are explored here.  In Sections~5 
a $\sigma$-$L$ relation is constructed and shown to be
equal to that obtained using SDSS early-type galaxy data, thereby laying to 
rest concerns that the
local sample of galaxies with direct SMBH mass measurements may be biased with
respect to the greater population (e.g.\ Yu \& Tremaine 2002; Bernardi et al.\ 2007). 
Furthermore, 
in section~6, the $K$-band $M_{\rm bh}$-$L$ relation and the barless 
$M_{\rm bh}$-$\sigma$ relation are shown to yield consistent results with 
neither giving SMBH masses greater than $\sim4 \times 10^9 M_{\sun}$.

\section{The ($M_{\rm bh}, \langle \mu \rangle _{\rm e}, R_{\rm e}$) plane}

Barway \& Kembhavi (2007, hereafter BK07) made the interesting claim that a
combination of two photometric parameters, namely the effective radius $R_{\rm
e}$ and the mean effective surface brightness $\langle \mu \rangle _{\rm e} =
-2.5\log\langle I \rangle _{\rm e}$, can be used to predict SMBH masses with a
greater degree of accuracy than single quantities such as luminosity or
velocity dispersion.
%

A tight relation exists between black hole mass, $M_{\rm bh}$, and the
luminosity, $L$, of the host bulge.  The luminosity can of course be expressed
in terms of two other
parameters because $L = 2 \pi R_{\rm e}^2 \langle I \rangle _{\rm e}$, where
$\langle I \rangle _{\rm e}$ is average intensity within the effective half
light radius, $R_{\rm e}$, of the bulge.  One question of interest is whether
the scatter about the $M_{\rm bh}$-$(R_{\rm e}^2 \langle I \rangle _{\rm e})$
relation can be reduced by allowing the exponents on $\langle I \rangle _{\rm
e}$ and $R_{\rm e}$ to deviate from their 1:2 ratio and, importantly, if this
results in less scatter than the other competing relations.
Given the small scatter about the Fundamental Plane ---
involving $R_{\rm e}$, $\langle \mu \rangle _{\rm e} = -2.5\log \langle I 
\rangle _{\rm e}$, 
and $\sigma$ --- and the tight relationship between 
$M_{\rm bh}$ and $\sigma$ (Merritt \& Ferrarese 2001;
Tremaine et al.\ 2002), one may indeed expect a well defined plane using the 
parameters $R_{\rm e}$, $\langle \mu \rangle _{\rm e}$, and $M_{\rm bh}$.

\subsection{Barway \& Kembhavi data} \label{Sec_BK1} 
 
This section examines BK07's claim that the total root mean square (r.m.s.) scatter 
in the $\log M_{\rm bh}$ direction, when using $\langle \mu \rangle _{\rm e}$ 
and $\log R_{\rm e}$ as predictor quantities of $M_{\rm bh}$, 
is 0.25 dex (and 0.19 dex when excluding the outlier NGC~4742). 

A simple linear, $Y=A+BX$, ordinary least squares regression analysis 
OLS$(Y$$\mid$$X)$ is performed with $Y = \log M_{\rm bh}$ and $X=\log
R_{\rm e} + b\langle \mu \rangle _{\rm e}$.  Solving for the parameters 
$A$, $B$, and $b$, 
this non-symmetrical regression gives the smallest r.m.s.\ residual in the $\log
M_{\rm bh}$ direction, which is what one wants when using the $M_{\rm
bh}$-$\langle \mu \rangle _{\rm e}$-$R_{\rm e}$ plane to predict $M_{\rm bh}$
in other galaxies (see Feigelson \& Babu 1992).  
The data for $\log M_{\rm bh}$, $\log R_{\rm e}$, and
$\langle \mu \rangle _{\rm e}$ have been taken from Table~1 in BK07.  Due to
the absence of reported errors on the quantities $\log R_{\rm e}$ and $\langle
\mu \rangle _{\rm e}$ in BK07, measurement errors are not included in the
regression, and subsequently no attempt to quantify the intrinsic scatter has
been made.
Parameter uncertainties are derived here using a bootstrap sampling of the data
points (i.e.\ sampling with replacement from the original sample) to produce
1000 Monte Carlo samples from which 1000 optimal fits are derived.  This
provides a histogram of each parameter from which one can compute the central
68.3\% width, which is used as the 1$\sigma$ uncertainty.

The optimal ($B$-band) solution using all 18 data points from BK07 is 
\begin{eqnarray}
\log(M_{\rm bh}/M_{\sun}) & = & (8.18\pm 0.09) + (3.15\pm 0.33)\log [R_{\rm e}/3\, 
                    {\rm kpc}] \nonumber \\
                & - & (0.90\pm 0.18)[\langle \mu \rangle _{\rm e,B} - 21.0].
\label{Eq_18}
\end{eqnarray}
The total scatter in the $\log M_{\rm bh}$ direction is 0.32 dex.  However the
total scatter in the $M_{\rm bh}$-$\sigma$ relation for this same galaxy set
is 0.31 dex.

BK07 performed an additional analysis, excluding 
NGC~4742 whose SMBH mass derivation has not yet appeared in a
refereed paper (see Tremaine et al.\ 2002) and may therefore potentially be
erroneous.  This galaxy also appeared as a clear outlier in their data. 
This does not necessarily mean the data point is in error; it may simply be 
a $3\sigma$ event, or the distribution of residuals may perhaps not be `normal'. 
Robust statistics requires that outlying data points not 
bias an analysis.  No single data point from a distribution 
should have the ability to significantly alter the result of an 
analysis dictated by the remaining population.  
The optimal relation after excluding NGC~4742 is given by the expression 
\begin{eqnarray}
\log(M_{\rm bh}/M_{\sun}) & = & (8.21\pm 0.07) 
   + (3.23\pm 0.26)[\log R_{\rm e}/3\, {\rm kpc}] \nonumber \\
      & - & (1.01\pm 0.13)[\langle \mu \rangle _{\rm e,B} - 21.0]
\label{Eq_murky}
\end{eqnarray}
with a total scatter of 
0.25 dex.
%
The low value of 0.19 dex reported by BK07 appears to have arisen by dividing
their scatter in the $\log R_{\rm e}$ direction (0.061) by the coefficient in
front of the $\log M_{\rm bh}$ term in their eq.3 (which is their fitted
plane).  However, this approach overlooks the three-dimensional nature of the plane and
consequently results in an over-estimation of the plane's ability to predict
black hole masses.  Computing the r.m.s.\ offset between the black hole masses
listed in Table~1 of BK07 (excluding NGC~4742) and the values predicted from
their plane (their eq.3), which can be re-written as $\log(M_{\rm bh}/M_{\sun}) = 8.28 +
3.13[\log R_{\rm e}/3\, {\rm kpc}] - 0.97[\langle \mu \rangle _{\rm e,B} -
21]$, one obtains a total scatter in the $\log M_{\rm bh}$ direction of 0.27
dex, not 0.19 dex, and greater than the value of 0.25 dex obtained above using 
Eq.\ref{Eq_murky}.

Although the claim in BK07 appears misplaced, based on an erroneous treatment
of the data, the idea tested there is a valid one.  In an effort to improve
the $M_{\rm bh}$-$\langle \mu \rangle _{\rm e}$-$R_{\rm e}$ plane's
reliability for predicting black hole masses, it is noted here that three of
the galaxies used by BK07 are known disk galaxies, or at least they are not
regular elliptical galaxies.  M32 may be a stripped S0 galaxy (Bekki et al.\
2001; Graham 2002), while NGC 2778 is a disk galaxy (Rix, Carollo \& Freeman
1999) as is NGC 4564 (Trujillo et al.\ 2004; see also figure~6 in Graham \&
Driver 2007a).  Consequently, the effective radii and mean surface
brightnesses which have been used for these three galaxies do not pertain to
their bulges.  IC~1459 is also excluded here due to the order of magnitude
uncertainty on its SMBH mass (Cappellari et al.\ 2002).  Excluding these four
galaxies plus NGC~4742 gives, from a reduced sample of only 13 galaxies, a
total scatter of 0.28 dex.  However, the total scatter in the $M_{\rm
bh}$-$\sigma$ relation for this cleaned galaxy set is 0.27 dex.  The $M_{\rm
bh}$-$\sigma$ relation therefore appears more competitive than the $M_{\rm
bh}$-$\langle \mu \rangle _{\rm e}$-$R_{\rm e}$ plane.

This sample size is obviously small and therefore makes it hard to reach
reliable conclusions.  The $M_{\rm bh}$-$\langle \mu \rangle _{\rm e}$-$R_{\rm
e}$ plane is thus investigated further with a larger galaxy sample in the
following subsection.

\subsection{Marconi \& Hunt data}\label{Sec_MH1}

Instead of using the $B$-band data in BK07, the scatter about the
$M_{\rm bh}$-$\langle \mu \rangle _{\rm e}$-$\log R_{\rm e}$ plane is explored
here using the
larger, homogeneous, $K$-band data set from Marconi \& Hunt's (2003; hereafter MH03) 27
``Group 1'' galaxies.  Using the minor-to-major axis ratio, $b/a$, of the
GALFITted (Peng et al.\ 2002) S\'ersic bulge component (Marconi \& Hunt,
priv.\ comm.), MH03's tabulated major-axis effective radii, $R_{\rm e,maj}$,
have been converted into a geometric mean radius $R_{\rm e}=\sqrt{R_{\rm
    e,maj}^2(b/a)}$ which is used here. 
%
While MH03 did not report any values for $\langle \mu \rangle _{\rm e}$, they
can be derived from the expression
\begin{equation}
\langle\mu\rangle_{\rm e} = m_{\rm tot} + 2.5\log[2\pi R_{\rm e,maj}^2(b/a)], 
\label{Eq_Mu}
\end{equation}
where $m_{\rm tot}$ is the apparent magnitude of the bulge (obtained from the absolute
magnitude and distance in Table~1 in MH03).
These values are shown in Table~\ref{Tab1}.


Four of the five galaxies which were excluded at the end of
Section~\ref{Sec_BK1} are in MH03's ``Group 1'' list.  They are again excluded
here for the same reasons.  Doing so, one obtains from the remaining 23
galaxies, using the $M_{\rm bh}$ values given in MH03,  
\begin{eqnarray}
\log(M_{\rm bh}/M_{\sun}) & = & (7.92\pm0.12) + (2.24\pm0.37)[\log R_{\rm e}/3\, {\rm
  kpc}] \nonumber \\
 & - & (0.54\pm0.14)[\langle \mu \rangle _{\rm e,K} - 17.5],
\label{Eq_MH20}
\end{eqnarray}
which has a total scatter in the $\log M_{\rm bh}$ direction of
0.33 dex.  The coefficients in Eq.\ref{Eq_MH20} 
are consistent with an $M_{\rm bh}$-$(I_{\rm
  e}R_{\rm e}^2)$ plane, and the total scatter in the $M_{\rm bh}$-$L$
  relation for these galaxies (0.34 dex) is comparable.
Moreover, the total scatter in the $M_{\rm bh}$-$\sigma$-$R_{\rm e}$ plane for
this same galaxy set is 0.28 dex.  It is therefore concluded that the $M_{\rm
bh}$-$\langle \mu \rangle _{\rm e}$-$\log R_{\rm e}$ plane is not warranted. 

The following section explores the $M_{\rm bh}$-$\sigma$-$R_{\rm e}$ plane,
and other planes involving $M_{\rm bh}$, $\sigma$ and some third parameter.

\section{The ($M_{\rm bh}$-$\sigma$)-$X$ plane}

MH03 explored the addition of $\log R_{\rm e}$ to the
$M_{\rm bh}$-$\sigma$ relation to create a ``fundamental plane for SMBHs''. 
From their 27 ``Group 1'' galaxies, 
they constructed a relation between $M_{\rm bh}$ and
$R_{\rm e}\sigma^2$ (proportional to the virial bulge mass), which 
resulted in an intrinsic dispersion\footnote{Intrinsic dispersion, 
sometimes called internal dispersion, is the scatter remaining after 
subtracting in quadrature, from the total scatter, 
the contribution from the assumed measurement uncertainties.}  
(total scatter) of 0.25 (0.30) dex in the $\log M_{\rm bh}$ direction.  
Allowing the exponents on the $R_{\rm e}$ and $\sigma$ terms to vary 
independently, Hopkins et al.\ (2007) used the same 27 Group 1 galaxies from
MH03 along with some updated measurements, to report that 
$\log(M_{\rm bh}/M_{\sun}) = (8.33\pm0.06) + (0.43\pm0.19)\log [R_{\rm e}/3\, {\rm kpc}] + 
(3.00\pm0.30)\log [\sigma/200\, {\rm km\, s}^{-1}], $
with an intrinsic scatter of 0.21 dex (and a total scatter of 0.30
dex, Hopkins 2007, priv.\ comm.).
%
%
For comparison, the $M_{\rm bh}$-$\sigma$ relation in Tremaine et al.\ (2002)
has an intrinsic (total) scatter of 0.27 (0.34) dex.  It therefore appears
that the introduction of a third parameter to the standard $M_{\rm bh}$-$\sigma$
relation may reduce the scatter and Hopkins et al.\ (2007) show that it does.
Here it is investigated which third parameter is optimal. 

From Graham \& Driver (2007a, hereafter GD07) the total scatter about the
log-quadratic M-n relation is reported to be 0.31 dex.  This highlights the
strong connection between $M_{\rm bh}$ and the radial structure in the stellar
distribution of the host bulge (see also Graham et al.\ 2007, their
Section~1), and hence the need to advance beyond $R^{1/4}$ models and their
associated luminosity/mass dependent biases (e.g., Trujillo et al.\ 2001; 
Brown et al.\ 2003).  This
Section explores whether the scatter about the $M_{\rm bh}$-$\sigma$ relation
is best reduced through the addition of the host bulge's $\log R_{\rm e}$,
$K$-band magnitude, or $\log n$.
The largest homogeneous sample of galaxies with published $M_{\rm bh}$ and $n$
values is that in GD07.  One of the strengths of the $M_{\rm bh}$-$n$ relation
is that photometrically uncalibrated images can be used. While this means that
GD07 do not have magnitudes, they do have bulge $R_{\rm e}$ values and
bulge-to-total ratios which can, when needed, be applied to the galaxy $M_K$
values in MH03.
The S\'ersic indices from GD07 pertain to the major-axis.  It is perhaps worth
noting that from very early on it was know that the major- and minor-axis need
not and do not have the same S\'ersic profile shape (Caon et al.\ 1993).  In
the presence of ellipticity gradients the S\'ersic index will vary with
position angle (Ferrari et al.\ 2004), and the value obtained from a
symmetrical 2D fit with a single S\'ersic index will match neither the major-
nor minor-axis value.  Moreover, the random viewing angles at which spheroids
are viewed will also introduce scatter to the $M_{\rm bh}$-$n$ relation (and
the $M_{\rm bh}$-$\sigma$ relation if the bulges are triaxial).

%
The S\'ersic indices and SMBH masses for the 27 galaxies tabulated in GD07 are
used here, along with the (geometric mean) effective radii (Table~\ref{Tab1}),
$K$-band magnitudes from MH03, and the central velocity dispersions from
Ferrarese \& Ford (2005, hereafter FF05).  NGC~6251 and NGC~7052 had a
different distance in GD07 and MH03, and have had their $R_{\rm e}$ and $M_K$
values adjusted to match the distances in GD07.
Although MH03 did not include/model NGC~1399 (Houghton et al.\ 2006), 
a velocity dispersion $\sigma = 344$ km s$^{-1}$ (HyperLeda) has been adopted, 
along with the S\'ersic index, $R_{\rm e}$ value, and $B$-band magnitude from 
D'Onofrio et al.\ (1994), adjusted to a distance of 20 Mpc (Tonry et al.\ 
2001), and using $b/a=0.94$ (NED) and $B-K = 4.14$ (Buzzoni 2005). 
The bulge parameters from Graham (2002) are used for NGC~221, 
along with a Johnson $R-K$ color of 2.34 (Buzzoni 2005).   
This left two galaxies (NGC~2778 and NGC~4564) which had $R_{\rm e}$ and
$M_K$ values pertaining to the galaxy rather than the bulge in MH03.
From the analysis in GD07, the values $R_{\rm e,maj}$ equals 0.25 and 0.31 kpc, 
and the $B/T$ ratios 0.21 and 0.24 have been adopted, respectively.

For these 27 galaxies one obtains an $M_{\rm bh}$-$\sigma$ relation similar to
that reported in Tremaine et al.\ (2002); it is such that
\begin{equation}
\log(M_{\rm bh}/M_{\sun})  = (8.09\pm0.07) + (4.08\pm0.40)\log [\sigma/200\, {\rm km\,
    s}^{-1}],
\label{Eq_MH_sig22}
\end{equation}
with $\Delta$, the total r.m.s.\ scatter in the $M_{\rm bh}$ direction, equal
to 0.31 dex.

The $M_{\rm bh}$-$\sigma$-$L$ plane for these galaxies is given by 
\begin{eqnarray}
\log(M_{\rm bh}/M_{\sun}) & = & (8.13\pm0.07) - (0.11\pm0.05)[M_K + 24] \nonumber \\
 & + & (3.34\pm0.48)\log [\sigma/200\, {\rm km\, s}^{-1}], 
 \Delta = 0.27\, {\rm dex}.
\end{eqnarray}
%
The $M_{\rm bh}$-$\sigma$-$R_{\rm e}$ plane is 
\begin{eqnarray}
\log(M_{\rm bh}/M_{\sun}) & = & (8.15\pm0.06) + (0.28\pm0.12)\log [R_{\rm e}/3\, {\rm kpc}]
\nonumber \\ 
 & + & (3.65\pm0.32)\log [\sigma/200\, {\rm km\, s}^{-1}],
  \Delta = 0.26\, {\rm dex},
\label{Eq_MH_MSR}
\end{eqnarray}
while the $M_{\rm bh}$-$\sigma$-$n$ plane is 
\begin{eqnarray}
\log(M_{\rm bh}/M_{\sun}) & = & (7.98\pm0.05) + (1.11\pm0.32)\log [n/3] 
\nonumber \\ 
 & + & (2.72\pm0.52)\log [\sigma/200\, {\rm km\, s}^{-1}],
  \Delta = 0.23\, {\rm dex}. 
\label{Eq_MH_MSN}
\end{eqnarray}

\begin{figure}
\includegraphics[angle=270,scale=1.0]{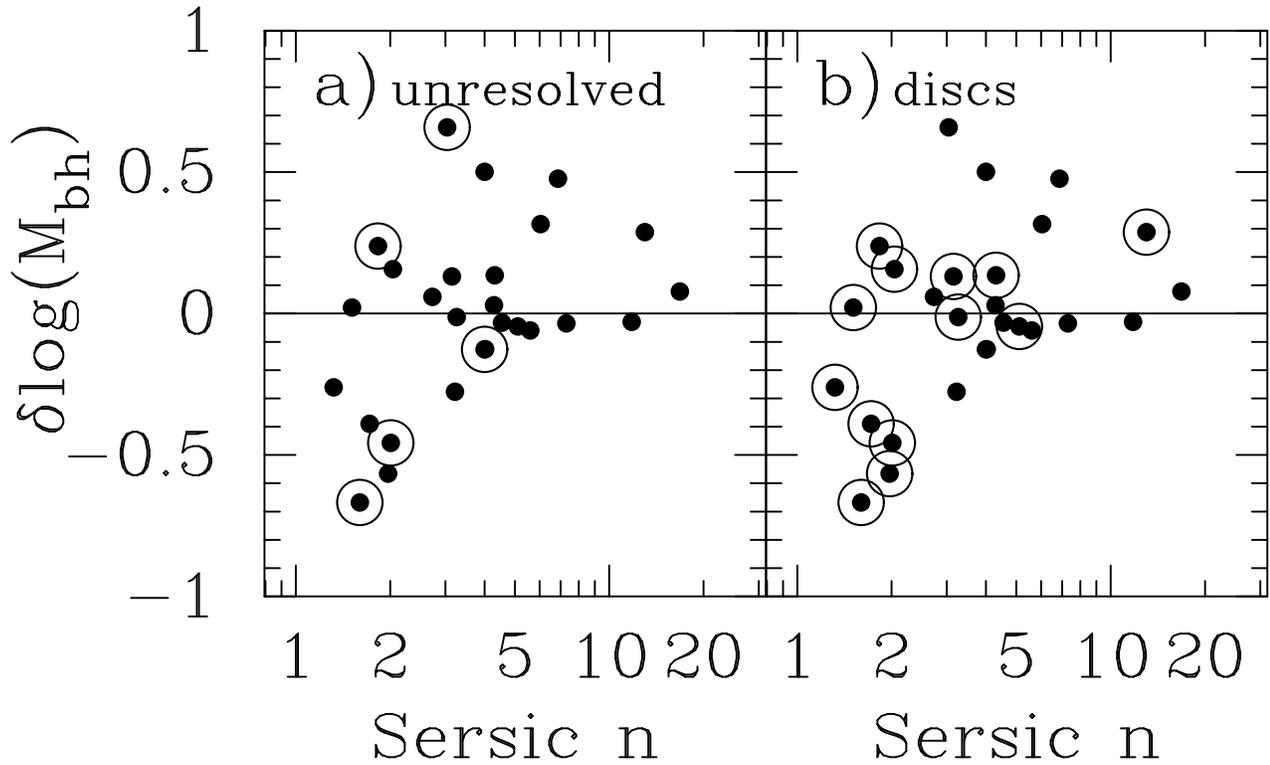}
\caption{
Using the 27 galaxies from GD07, the residuals about their $M_{\rm
bh}$-$\sigma$ relation (constructed to minimize the scatter in the $\log
M_{\rm bh}$ direction) are shown against their S\'ersic index $n$.  Panel a)
highlights the 5 galaxies with an unresolved SMBH sphere-of-influence
(Table~II in FF05) --- they are not responsible for the trend. 
Panel b) shows the 13 disk galaxies (circled); they are responsible
for the trend seen in this figure. 
}
\label{Fig_MSN}
\end{figure}

Given the best performer is the $M_{\rm bh}$-$\sigma$-$n$ plane, the residuals
about the $M_{\rm bh}$-$\sigma$ relation are plotted in Fig.\ref{Fig_MSN}
against the bulge S\'ersic index.  From Fig.\ref{Fig_MSN}a it is clear that a
trend will still persist after the exclusion of the five galaxies whose SMBH
sphere-of-influence is not resolved (according to Table~II from FF05).  On the
other hand, Fig.\ref{Fig_MSN}b reveals that the trend is caused by (some of)
the disk galaxies.  This intriguing aspect is explored further in the
following section.

\section{The ``barless $M_{\rm bh}$-$\sigma$'' relation}

\subsection{Graham \& Driver data}

Looking at Fig.\ref{Fig_MSN}, much of the trend is due to some
five data points from the small bulges of disk galaxies.
While these five bulges have small ($R_{\rm e} < 1$ kpc)  effective 
radii, some of the other disc galaxies have comparable radii but do not
deviate from the $M_{\rm bh}$-$\sigma$ relation.  
These five systems have SMBH masses $\sim$0.5 dex below the best fitting $M_{\rm
  bh}$-$\sigma$ relation. 
Intriguingly, all of these five disk galaxies
have been identified in the literature as containing bars.  They are: the 
Milky Way (e.g., L{\'o}pez-Corredoira et al.\ 2007, their Fig.5); 
NGC~1023 (Debattista et al.\ 2002, their Fig.~4b; Sil'chenko 1999); 
NGC~2778 (weak bar peaks at $5\arcsec$; Rest et al.\ 2001, their Fig.~8; Trujillo et al.\ 2004); 
NGC~2787 (Erwin et al.\ 2003, their Fig.~1); 
and NGC~3384 (Busarello et al.\ 1996, their Fig.~7; Cappellari \& Emsellem
2004; Erwin 2004; Meusinger \& Ismail 2007). 
In sharp contrast to this, only one of the other 8 disk galaxies (NGC~4258,
van Albada 1980) is
classified in NED as having a bar; therefore, if any of the other seven disk galaxies do
possess a bar, it must be weak.  
If the probability of a disk galaxy having a bar is equal to the probability
of not having a bar, 
then the distribution in Figure~\ref{Fig_MSN} has a 1 in 1024 chance of occurring. 
If 75\% of disk galaxies have bars (e.g., Eskridge et al.\ 2000;  
see also Knapen et al.\ 2000 and Marinova \& Jogee 2007), then 
the observed distribution has less than a one in ten thousand likelihood of
occurring by chance.

\begin{figure}
\includegraphics[angle=270,scale=0.7]{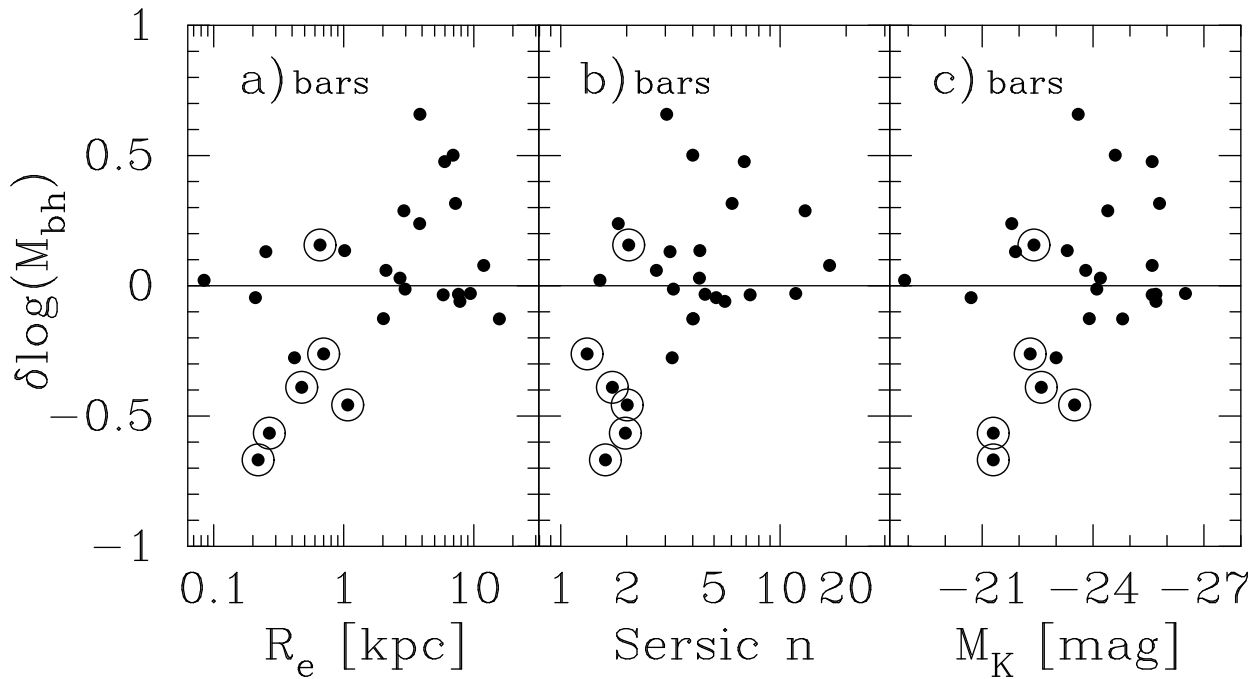}
\caption{
Similar to Fig.~\ref{Fig_MSN} except that the $\delta \log M_{\rm bh}$ values
are shown against a) the geometric mean effective radii from Table~1, b) the
S\'ersic indices $n$ from GD07, and c) the absolute $K$-band magnitudes $M_K$
from MH03.  The barred galaxies have been circled; they are responsible for
the bulk of the apparent trends. 
%
}
\label{Fig_RNM}
\end{figure}

In Fig.\ref{Fig_RNM}, the barred galaxies can be seen to be largely 
responsible for the reduced scatter when going from the $M_{\rm bh}$-$\sigma$
relation to the 
$M_{\rm bh}$-$\sigma$-$R_{\rm e}$, 
$M_{\rm bh}$-$\sigma$-$L$,  and 
$M_{\rm bh}$-$\sigma$-$n$ planes.  
In other words, these galaxies deviate from the $M_{\rm bh}$-$\sigma$
relation.  
%
It is therefore of interest to re-derive the $M_{\rm
bh}$-$\sigma$ relation excluding those galaxies with bars. 
For the 21 non-barred galaxies in GD07 
\begin{equation}
\log(M_{\rm bh}/M_{\sun})  = (8.20\pm0.05) + (3.83\pm0.10)\log [\sigma/200\, {\rm km\,
    s}^{-1}],
\label{Eq_cool}
\end{equation}
with a total scatter of 
0.22 dex (cf.\ Eq.\ref{Eq_MH_sig22}). 
%
Aside from NGC~4258, the barred galaxies have an offset in the 
$\log M_{\rm bh}$ direction of 0.5 to 0.8 dex.

In passing it is noted that a prolate bulge, even in the absence of a bar,
will have a smaller effective radius when viewed along its major-axis (e.g.,
Lanzoni \& Ciotti 2003).  A more detailed investigation of the above galaxies
could therefore include how the measured size, magnitude, and concentration of
the spheroidal component, and the velocity dispersion, changes with the
orientation of the bulge and bar.

Although Figures~\ref{Fig_RNM} 
might appear to hint that the barred galaxies have smaller effective radii
than the non-barred disk galaxies, a KS test reveals no significance (at even
the 1$\sigma$ level) that the cumulative distribution function for the barred
and unbarred disk galaxies' effective radii may be different.  Similarly,
Student's t-test reveals no significant difference between the means of each
distribution, with only an 86\% ($<$1.5$\sigma$) probability of difference. 

It is pertinent to ask whether the inclusion of an additional parameter to the
above ``barless $M_{\rm bh}$-$\sigma$'' relation is warranted.  The answer
appears to be `no'.  Reductions of not more than 0.01 dex are achieved through the
addition of either $R_{\rm e}$, $L$, or $n$.  This implies that a fundamental
plane for SMBHs is not appropriate; if it was, it should equally apply to
galaxies with and without bars. 
 
%
%

A bar may result in the fueling of the SMBH (e.g., Wyse 2004;
Ohta et al.\ 2007), perhaps eventually bringing its mass in line with the
$M_{\rm bh}$-$\sigma$ relation.  Although, the large ratio of barred galaxies
to active galaxies in the Universe today would seem to argue against this as a
common phenomenon, as does the incidence of bars in Seyfert and normal disk
galaxies (Mulchaey \& Regan 1997; Ho et al.\ 1997; although see Crenshaw et
al.\ 2003).  It may however be that some other physical property 
such as nuclear disks or kinematically decoupled cores are influencing the 
measured velocity dispersions. 

\begin{figure}
\includegraphics[angle=270,scale=0.7]{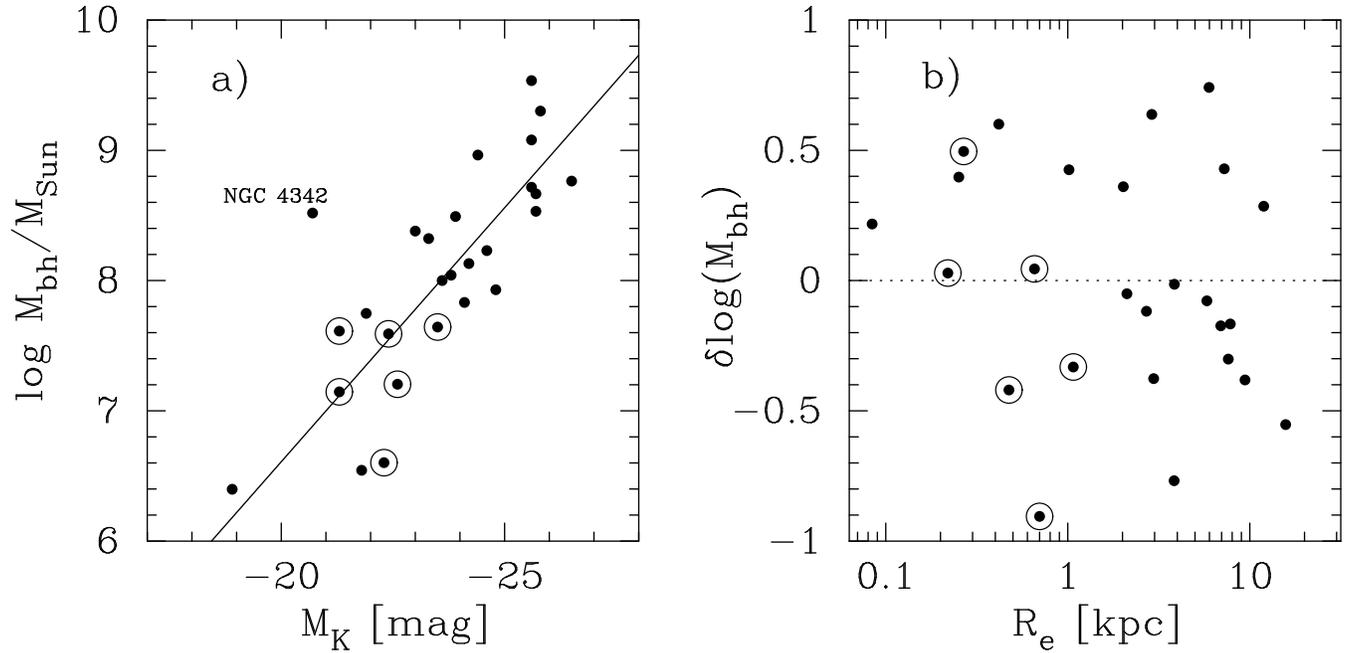}
\caption{
Panel a) $M_{\rm bh}$-$L$ relation for the 27 galaxies from GD07.
Panel b) shows the residuals versus their geometric mean effective radii. 
The six barred galaxies have been circled. 
}
\label{Fig_MLR}
\end{figure}

Bar instabilities are believed to lead to the formation of pseudobulges.  Such
evolution may have resulted in (pseudo)bulges with an increased velocity
dispersion and luminosity but a relatively anaemic SMBH (unless it also grew
during the formation of the pseudobulge).  If the barred galaxies do indeed
have discrepantly low SMBH masses, rather than high $\sigma$ values, they
should also appear as systematic outliers in the $M_{\rm bh}$-$L$ diagram.
Figure~\ref{Fig_MLR} reveals that this is not the case, suggesting that the
SMBH masses are okay but the velocity dispersions are discrepant.

Perhaps the bar dynamics have biased the measurement of the bulge velocity
dispersion.  The non-circular (streaming) motions of stars in bars obviously
deviates from that of the random motions of the bulge stars and may
potentially interfere with the central velocity dispersion measurements.  An
alignment of the (radial orbits in the) bar with our line-of-sight may result
in such a scenario\footnote{Velocity dispersion drops (M\'arquez et al.\ 2003;
  Wozniak et al.\ 2003) would, however, work in the opposite sense.}.  Indeed,
the barred galaxy NGC~3384 is highly inclined and has the bar closely aligned
with the projected minor axis (Busarello et al.\ 1996; Erwin et al.\ 2004).
In the case of NGC~1023, a quick visual inspection reveals a disk/bar position
angle of 80$^{\circ}$/72$^{\circ}$, suggesting that the inclination of this
galaxy does not result in us looking down the length of the bar.  However, as
Debattista (2002) revealed, after deprojecting this galaxy, it was found to
have a strong bar whose position angle is 102$^{\circ}$ offset from the
galaxy's (projected) major-axis.  That is, we {\it are} in fact looking down
the barrel of the bar in this galaxy.
A full treatment of each of the barred galaxies is beyond the scope of this
paper, it is however noted that the (projected) bar position angles
can appear more aligned with the (projected) major-axis 
than they are in reality, as is the case with NGC~1023.

\subsection{Tremaine et al.\ data}\label{Sec_Tre}

Removing the seven barred galaxies (the six mentioned above plus NGC~4596)
from the Tremaine et al.\ (2002) sample of 31 galaxies (with the SMBH mass for
NGC~821 updated with the value in Richstone et al.\ 2007), application of
Tremaine et al.'s regression technique gives
\begin{equation}
\log(M_{\rm bh}/M_{\sun})  = (8.21\pm0.05) + (3.89\pm0.26)\log [\sigma/200\, {\rm km\,
    s}^{-1}],
\label{Eq_Trem}
\end{equation}
with an intrinsic scatter of 0.17 dex (cf.\ 0.27 dex using the original 31
galaxies) and a total scatter of 0.25 dex (cf.\ 0.34 dex using the original
31 galaxies). 
This is in agreement with Eq.\ref{Eq_cool} which used a slightly different
sample and velocity dispersions from FF05.

Construction of a barless $M_{\rm bh}$-$\sigma$-$M_B$ plane, with the absolute
$B$-band magnitude, $M_B$, taken
from Tremaine et al.\ (2002), has the same scatter as the barless $M_{\rm
bh}$-$\sigma$ relation.  Furthermore, the intrinsic scatter from Eq.\ref{Eq_Trem} is
smaller than the value of 0.21 dex reported in Hopkins et al.\ (2007).
It seems reasonable to conclude that previous claims
for the existence of ``fundamental planes for SMBHs'' have been influenced by
the presence of barred galaxies. 

Using the Tremaine et al.\ (2002) sample and performing a regression which
minimizes the residuals in the $\log \sigma$ direction, rather than the $\log
M_{\rm bh}$ direction, the intercept and slope of the barless $M_{\rm
bh}$-$\sigma$ relation are 8.21 and 4.05.  A symmetrical regression 
will therefore have a slope around $(3.89 + 4.05)/2 = 3.97$.
Using the symmetrical bisector linear regression routine BCES from Akritas \&
Bershady (1996), one obtains 
\begin{equation}
\log(M_{\rm bh}/M_{\sun})  = (8.20\pm0.05) + (3.95\pm0.25)\log [\sigma/200\, {\rm km\,
    s}^{-1}]. 
\label{Eq_BCES}
\end{equation}
This expression should be preferred when trying to understand the physical
processes responsible for the correlation (see Feigelson \& Babu 1992),
although this point is perhaps moot given Eq.\ref{Eq_BCES}'s consistency with
Eq.\ref{Eq_cool} and \ref{Eq_Trem}.

\subsection{Marconi \& Hunt data}\label{Sec_MH2}

The analysis in Hopkins et al.\ used the 27 ``Group 1'' galaxies from MH03. 
In an effort to better understand the result in Hopkins et al., the original
data from MH03 is analyzed here.  Given that MH03's 
$M_{\rm bh}$ and $\sigma$ values for the Milky Way and M31 were not in dispute
(only their $K$-band magnitudes were somewhat in doubt), these two galaxies
have been included here with the 27 ``Group 1'' galaxies.

\begin{figure}
\includegraphics[angle=270,scale=0.7]{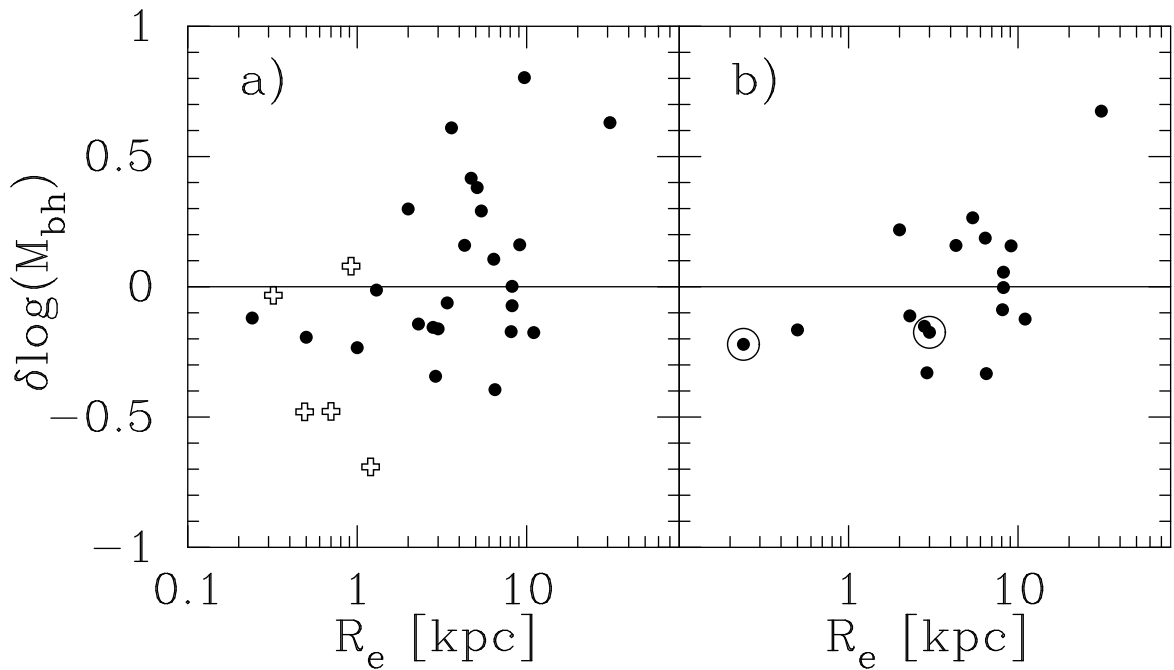}
\caption{
Panel a) Residuals about the $M_{\rm bh}$-$\sigma$ relation
using the 27 `Group 1' galaxies from MH03 plus their data for the Milky Way and
M31 are shown against their tabulated major-axis effective radii. 
The five barred galaxies are denoted with crosses.  Panel b) Residuals
about the $M_{\rm bh}$-$\sigma$ relation when using only the elliptical
galaxies (according to MH03) to construct the $M_{\rm bh}$-$\sigma$ relation.
The non-elliptical galaxies NGC~221 and NGC~4564 have been circled.  The point in
the top right is Cygnus~A at 240 Mpc (using $M_{\rm bh} = 2.9\times 10^9
M_{\sun}$). 
}
\label{Fig3}
\end{figure}

The residuals, in the $\log M_{\rm bh}$ direction, 
about the $M_{\rm bh}$-$\sigma$ relation are shown in
Figure~\ref{Fig3}a for the above 29 galaxies.  
The trend between the residuals 
and the effective radii of the host spheroids does indeed appear to suggest
the need for a fundamental plane type relation, akin to that proposed by 
MH03, and later by Hopkins et al.\ (2007) 
and also Aller \& Richstone (2007) using a sample of 23 galaxies. 
However, once one identifies the  
(five) barred galaxies in the above sample, the evidence for such a plane is 
reduced.  The three galaxies with the largest negative residual are barred
galaxies. 

The three galaxies with the highest positive residuals in Figure~\ref{Fig3}a --- 
two of which still seem to advocate the need for a `fundamental plane' --- 
are, in order of increasing $R_{\rm e}$: the radio galaxy Centaurus A, the Seyfert galaxy
NGC~5252 at $\sim$100 Mpc, and Cygnus A at a distance of 240 Mpc. 
The SMBH mass estimate for Centaurus A that was used by MH03 and used in
Figure~\ref{Fig3} has however since been revised downward
by more than a factor of two (Marconi et al.\ 2006) 
and so Cen~A is therefore now known not to be an outlier. 
Due to their distances\footnote{The third and 
only other galaxy further than 35 Mpc is NGC~6251 at 107 Mpc.} and somewhat 
disturbed morphology, none of 
these galaxies had been fitted with a S\'ersic profile by GD07, 
nor had they been included in Tremaine et al.\ (2002).  
While one can conclude that (some) barred galaxies deviate significantly from
the $M_{\rm bh}$- $\sigma$ relation, the inclusion of NGC~5252 and Cygnus A
may present some evidence in favor of a fundamental plane for SMBHs. 

If a fundamental plane for black holes does exist, demonstrating its existence
with an elliptical only sample would help eliminate concerns that unrelated
processes pertaining to bars and disks are misleading us.  Figure~\ref{Fig3}b
shows the residuals about the $M_{\rm bh}$- $\sigma$ relation for the 17
elliptical galaxies from the sample of 29.  One can immediately see that there
is, as yet, no convincing evidence for an elliptical galaxy SMBH fundamental
plane involving $M_{\rm bh}$, $\sigma$, and $R_{\rm e}$.  Given the obvious
need for more data, the following section introduces and uses new SMBH data
obtained after 2003.

\subsection{Additional data \label{Sec_Add}}

Since MH03's paper, additional galaxies have had their SMBH
masses measured.  These are provided in Table~\ref{Tab3}, along with galaxies
from MH03 for which some updates have become available, giving a
total of 40 galaxies with direct SMBH mass measurements. 
An additional 15 galaxies with somewhat uncertain SMBH mass estimates (see
FF05) are listed in Table~\ref{Tab4}.  Although these are
not used here, they are provided for a sense of awareness as to further
galaxies which may be useful in the future.

\begin{figure}
\includegraphics[angle=270,scale=1.0]{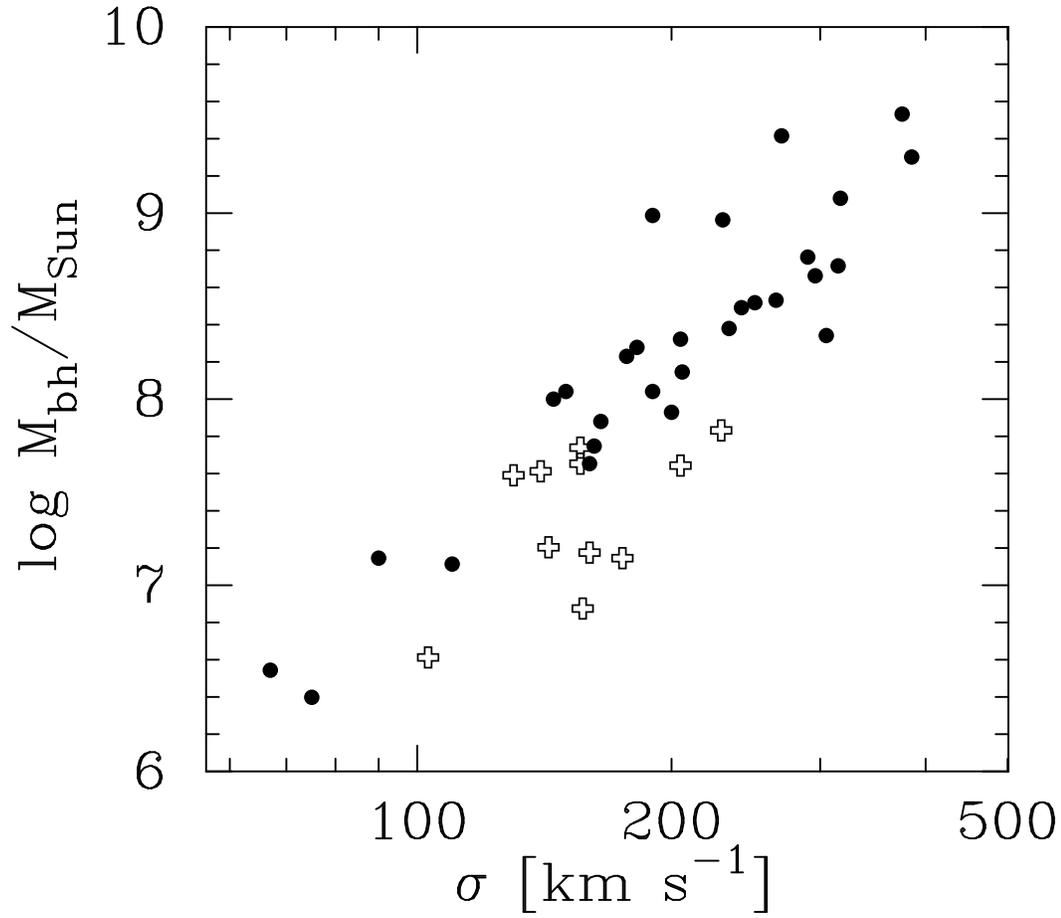}
\caption{
$M_{\rm bh}$-$\sigma$ diagram for 40 galaxies 
(see section~\ref{Sec_Add}). 
The 11 barred galaxies are denoted with a cross. 
}
\label{Fig4}
\end{figure}

Using (i) the (updated) data for the 27 `Group 1' galaxies from MH03 
(except for IC~1459 and NGC~4594, whose $M_{\rm bh}$ values are somewhat
uncertain), plus (ii) MH03's ten `Group 2' galaxies (minus
NGC~1068, NGC~4459, and NGC~4596 for which the SMBH mass estimates are also
not secure), plus (iii) the nine new galaxies in Table~\ref{Tab3} (excluding
NGC~2748 for which there is no published velocity dispersion), gives a total
sample of (25+7+8=) 40 galaxies from which an updated $M_{\rm bh}$-$\sigma$
diagram has been constructed (Figure~\ref{Fig4}).
%
For these galaxies, the $M_{\rm bh}$-$\sigma$ relation and the $M_{\rm
bh}$-$\sigma$-$R_{\rm e}$ plane are given in Table~\ref{Tab5}, along with the
associated total scatter.  For the full data set, one obtains an $M_{\rm
bh}$-$\sigma$ relation in good agreement with Tremaine et al.\ (2002).  One
also has $M_{\rm bh} \propto \sigma^{3.23\pm0.28} R_{\rm
e}^{0.43\pm0.11}$, in agreement with the result in Hopkins et al.\ (2007).

\begin{figure}
\includegraphics[angle=270,scale=0.7]{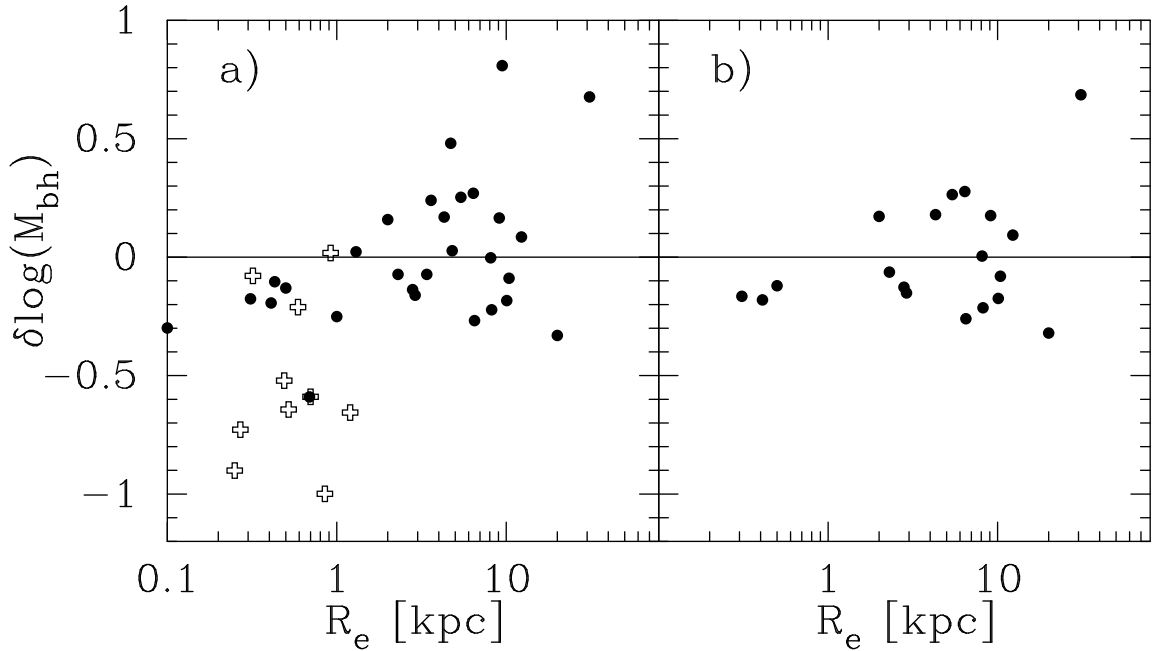}
\caption{
Panel a) Residuals about the $M_{\rm bh}$-$\sigma$ relation constructed using
the 29 non-barred galaxies (see section~\ref{Sec_Add}).  
The residual offset of the ten barred galaxies
that have $R_{\rm e}$ values are denoted with a cross.  Panel b) Residuals
about the $M_{\rm bh}$-$\sigma$ relation constructed using the 19 elliptical
galaxies.  
}
\label{Fig5}
\end{figure}

However, from Fig.~\ref{Fig4} one can clearly see that many of the barred galaxies
deviate from the $M_{\rm bh}$-$\sigma$ relation and are obviously responsible
for some of the perceived need for a fundamental plane.  If these galaxies had
$R_{\rm e}$ values that were smaller than any of the other bulges, then one
could argue that the presence of the bar may have nothing to do with their
displacement from the $M_{\rm bh}$-$\sigma$ relation, and that a `fundamental
plane' is needed.  However this is not the case, that is, other small
spheroids exist which do not deviate from the $M_{\rm bh}$-$\sigma$ 
relation.  The only non-barred galaxy with a notable negative $\delta \log
M_{\rm bh}$ residual in Figure~\ref{Fig4} and \ref{Fig5} is the LINER galaxy NGC~3998 
(De Francesco et al.\ 2006).
As remarked by Fisher (1997), this galaxy has a very steep central velocity
dispersion profile, dropping from $\sim$320 km s$^{-1}$ at $r=0$ to $\sim$160
km s$^{-1}$ at $r=4$ arcseconds (270 pc)
A velocity dispersion of 210 (or 250) km s$^{-1}$ for this galaxy would 
result in a zero (or 1$\sigma$) residual about the $M_{\rm bh}$-$\sigma$
relation. 
%
%
%

Removing the 11 barred galaxies from the sample of 40, 
one obtains the ``barless'' $M_{\rm bh}$-$\sigma$ relation given in Table~\ref{Tab5}. 
The vertical residuals about this relation are shown in Figure~\ref{Fig5}a,
along with the offsets of the barred galaxies relative to this $M_{\rm
  bh}$-$\sigma$ relation defined by the non-barred galaxies. 
While seven of the ten barred galaxies with known $R_{\rm e}$ values are
responsible for much of the trend between the $M_{\rm bh}$-$\sigma$ residuals
and $R_{\rm e}$, the non-barred galaxies do still reveal a trend. 
Indeed, from Table~\ref{Tab5}, one can see that the non-barred galaxies 
favor a fundamental plane relation. 
However, it is noted that 
removal of just two galaxies (NGC~3998 and Cygnus A) from the sample of 29
non-barred galaxies 
leaves the coefficient in front of the $\log(R_{\rm e}/3)$
term inconsistent with a value of zero at a significance of less than 
2$\sigma$.  It is disconcerting that just a couple of points are responsible
for the apparent plane. 
The bulk of the data does not suggest the need for a fundamental plane.

As noted previously, 
to be certain that a 3-parameter fundamental plane is required to describe the
connection between SMBHs and their host spheroids, one would ideally like to
use a sample of elliptical galaxies.  This would ensure that the `plane' is
not a byproduct of additional physical mechanisms or biases related to the
presence of a disk and/or bar.  Using the 19 elliptical galaxies from the 
sample of 40 galaxies, one has 
\begin{equation}
\log(M_{\rm bh}/M_{\sun})  = (8.25\pm0.05) + (3.68\pm0.25)\log [\sigma/200\, 
   {\rm km\, s}^{-1}],
\label{Eq_fin}
\end{equation}
with a total scatter of 0.24 dex.  Exclusion of the single data point for
Cygnus~A, the galaxy with the greatest residual offset in Fig.\ref{Fig5}b,
reduces the total scatter to 0.18 dex.  This is the same scatter as that about
the best fitting $M_{\rm bh}$-$\sigma$-$R_{\rm e}$ plane to this set of 18
elliptical galaxies.
The elliptical galaxies therefore do not provide substantial support for the
existence of an $M_{\rm bh}$-$\sigma$-$R_{\rm e}$ fundamental plane for SMBHs
(see Table~\ref{Tab5}).  When using all 19 elliptical galaxies, 
the 2$\sigma$ uncertainty on the coefficient in front
of the $R_{\rm e}$ term ranges from -0.08 to 0.55.  This parameter is
inconsistent with a value of zero at only the 1.4$\sigma$ level.  Moreover, 
removing just one data point (Cygnus~A) reduces the coefficient in front of 
the $R_{\rm e}$ term to $0.09\pm0.11$. 

Given the small sample sizes involved, it may be premature to completely rule
out the existence of a fundamental plane for SMBHs.  Some may object to the
removal of outlying data points, which is why equations using both complete
and adjusted data sets have been provided.  Most would however acknowledge that 
a certain degree of caution must be associated with any conclusion that hinges
on outlying data points.  Indeed, for similar reasons, $3\sigma$ clipping of 
distributions is a somewhat common practice these days. 
One thing which is clear is that the biasing presence of disc (especially
barred) galaxies appear responsible for much of the alleged evidence for 
requiring a fundamental plane for SMBHs. 

One should not use either the ``barless $M_{\rm bh}$-$\sigma$''
relation nor the standard $M_{\rm bh}$-$\sigma$ relation for barred galaxies
because the resultant SMBH mass estimates may be in error (too high) by 0.5 to
1.0 dex.  One should also not apply an $M_{\rm bh}$-$\sigma$-$R_{\rm e}$ fundamental
plane in the hope of accounting for barred galaxies because such a plane will
introduce a bias to the non-barred galaxies.

\section{The $\sigma$-$L$ relation and sample bias}

There has been some concern recently that the $M_{\rm bh}$-$\sigma$ and/or
$M_{\rm bh}$-$L$ relation may be biased, and that they are not consistent with
each other.  
Lauer et al.\ (2007), Bernardi et al.\ (2007), and Graham (2007, his
Appendix~A) have reported a slight difference in the $\sigma$-$L$ relation
between the local sample of galaxies with direct SMBH masses and the greater
population.  If correct, this implies that either the $M_{\rm bh}$-$\sigma$ or
$M_{\rm bh}$-$L$ relation may be biased.  Given the offset nature of some of
the barred galaxies in the $M_{\rm bh}$-$\sigma$ diagram --- in the sense that
they have overly large velocity dispersions for their SMBH masses --- it is
apposite to explore if the barred galaxies may be responsible for the
allegedly biased nature of these local inactive galaxy samples.

\begin{figure}
\includegraphics[angle=270,scale=0.7]{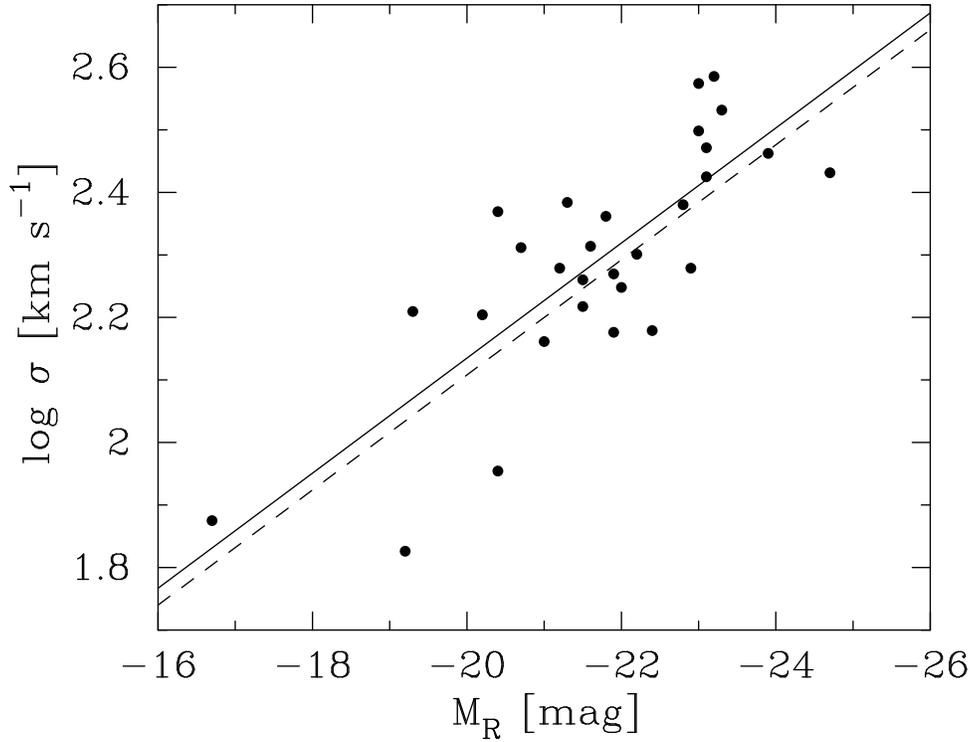}
\caption{
The $R_c$-band $\sigma$-$L$ relation for SDSS early-type galaxies (dashed line) taken
  from Tundo et al.\ (2007, their Eq.4), and for non-barred galaxies with
  direct SMBH mass estimates in MH03 (solid line, Eq.\ref{Eq_bias}).  
  The outlying galaxy NGC~4342 has been excluded. 
}
\label{Fig6}
\end{figure}

The Group 1 and 2 galaxy data from MH03 is used here, along with the updates
noted in Table~\ref{Tab3}.  The 7 barred galaxies from MH03's 37 `Group 1' and
`Group 2' galaxies are excluded, and the $K$-band magnitudes have been
converted to the $R_c$-band using $R_c-K=2.6$ (Buzzoni 2005).  An uncertainty
of 0.3 mag and 5\% is assigned to the magnitudes and velocity dispersions,
respectively.  Applying the regression analysis scheme from Tremaine et al.\
(2002) to minimize the scatter in the $\log \sigma$ direction, the optimal
$\sigma$-$L$ relation is
\begin{equation}
\log \sigma = 2.23\pm0.03 - (0.092^{+0.018}_{-0.012})[M_R +21], 
\label{Eq_bias}
\end{equation}
which is shown in Figure~\ref{Fig6}.  The parameter uncertainties have been 
estimated from a Monte Carlo bootstrap analysis. 
Although MH03 note that the $M_K$ value for M31 may be in error, excluding it
from the regression has no effect on Eq.\ref{Eq_bias}.  However, 
the extreme outlying point NGC~4342, the smallest and faintest spheroid
from MH03's sample after M32, is excluded from this regression. 

The reason for constructing an $R_c$-band relation was to allow a comparison
with the result from Tundo et al.\ (2007, their Eq.4), which is a SDSS
$r^{\prime}$-band $\sigma$-$L$ relation for early-type SDSS galaxies, the
majority of which presumably do not have bars.  Using $r^{\prime}-R_c = 0.24$
(Fukugita et al.\ 1995), Tundo et al.'s expression is such that $\log \sigma =
0.27 - 0.092M_{R_c} = 2.20 - 0.092(M_{R_c} + 21)$, in remarkable agreement
with Eq.\ref{Eq_bias}.  Therefore, it is not yet established that the local
sample of galaxies with direct SMBH mass measurements is biased.

\section{$M_{\rm bh}$-$\sigma$ versus $M_{\rm bh}$-$L$ \label{Sec_Max}}

Given that the local (predominantly inactive) sample of galaxies with direct
SMBH mass measurements appears to be unbiased with respect to the greater
population, it is appropriate to re-examine whether the (barless) 
$M_{\rm bh}$-$\sigma$ and $M_{\rm bh}$-$L$ relations predict different SMBH
masses.  Indeed, 
it has been claimed that these 
relations are not consistent with each other, in the sense that massive
galaxies are predicted to have more massive SMBHs when using the $M_{\rm
bh}$-$L$ relation, with values up to $10^{10} M_{\sun}$ (Lauer et al.\ 2007).

This discrepancy is investigated here by first looking at the upper extremity
of the $M_{\rm bh}$-$\sigma$ relation.  At 400 km s$^{-1}$, it turns out that
both the old $M_{\rm bh}$-$\sigma$ relation --- as given by Tremaine et al.'s
(2002) regression of $\log M_{\rm bh}$ on $\log \sigma$ --- and the new
relation (Eq.\ref{Eq_fin}) predict the same black hole mass:
$2.3^{+1.7}_{-1.0}\times 10^9 M_{\sun}$.  A value of 400 km s$^{-1}$ is used
here due to the rapid decline in the number density of systems with higher
velocity dispersions (Sheth et al.\ 2003; Bernardi et al.\ 2006).  This upper
black hole mass agrees well with that from the $M_{\rm bh}$-$n$ relation in
GD07, where $M_{\rm bh,upper} = 1.2^{+2.6}_{-0.4}\times 10^9 M_{\sun}$,
implying an upper ($1\sigma$) SMBH mass limit of $\sim 4\times 10^9 M_{\sun}$.

For the ($K$-band) $M_{\rm bh}$-$L$ relation, $ \log(M_{\rm bh}/M_{\sun}) =
(8.29\pm0.08) - (0.37\pm0.04)[M_K+24]$ (Graham 2007), to predict a more
massive black hole than $2.3\times 10^9 M_{\sun}$ requires a spheroid with
$M_K < -26.9$ mag.  From the $K$-band magnitudes for 102 brightest cluster
galaxies (BCGs) in Stott et al.\ (2007), while many galaxies are close to this
limit, only two are brighter (after the small adjustment of 0.1 mag when
switching from $H_0 = 70$ to 73 km s Mpc$^{-1}$).  Furthermore, from the
(corrected) SDSS $r^{\prime}$-band BCG magnitudes in both Desroches et al.\
2007, their Figure~9) and Liu et al.\ (2008, their Figure~13) we see that the
brightest magnitudes truncate at $M_{r^{\prime}} \sim -24.2$ mag.  Using
$r^{\prime} - K = 2.8$, this corresponds to a $K$-band magnitude of $-27.0$
mag and an ($M_{\rm bh}$-$L$)-derived SMBH mass of $2.5\times 10^9 M_{\sun}$.
It therefore appears that the $M_{\rm bh}$-$L$ relation 
does not predict higher SMBH masses than the $M_{\rm bh}$-$\sigma$ relation. 
The near-infrared analysis by Batcheldor et al.\ (2007) also supports this
picture (but see Lauer et al.\ 2007). 
The $M_{\rm bh}$-$M_{\rm bh}$ diagram from Lauer et al.\ (2007, their Fig.2)
is at odds with the above result.  It is however first noted that no non-BCG
in Lauer et al.\ has a magnitude brighter than NGC~6876 at $M_V = -23.49$ mag
($H_0 = 73$).  Assuming a $V-K$ color of 3.22 for elliptical galaxies (Buzzoni
2005), this magnitude corresponds to $M_K = -26.71$ mag, giving a black hole
mass of $2.0\times 10^9 M_{\sun}$, consistent with the upper bound from the
$M_{\rm bh}$-$\sigma$ relation.

\begin{figure}
\includegraphics[angle=270,scale=0.7]{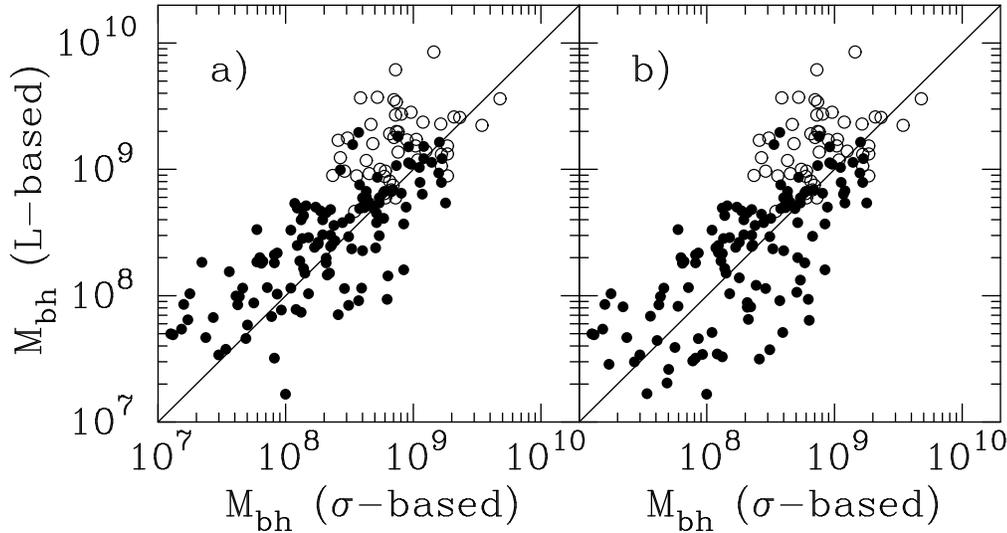}
\caption{
$M_{\rm bh}$ masses for the galaxies tabulated in Lauer et al.\ (2007) 
  obtained using the new $M_{\rm bh}$-$\sigma$ relation ((Eq.\ref{Eq_fin})
and the $M_{\rm bh}$-$L$ relation from Graham (2007, his Eq.14). 
In panel b) the $R^{1/4}$ bulge magnitudes of the disc galaxies have been adjusted
as described in Section~\ref{Sec_Max}.
The BCGs and normal galaxies are denoted with open circles and dots,
respectively. 
}
\label{Fig7}
\end{figure}

To try and resolve the issue with the BCGs in Lauer et al., their $M_{\rm
bh}$-$M_{\rm bh}$ diagram is reproduced here after applying a number of
updates.  First, the new $M_{\rm bh}$-$\sigma$ relation (Eq.\ref{Eq_fin}) is
applied to the velocity dispersions tabulated in Lauer et al\footnote{While this is
not ideal because galaxies with bars may have their SMBH masses
over-estimated, at the high mass end we do not predict larger SMBH masses than the $M_{\rm
  bh}$-$\sigma$ relation from Tremaine et al.\ (2002) on this same data.}.  
Second, the above mentioned $K$-band $M_{\rm bh}$-$L$
relation is used (and $V-K = 3.22$ applied).  As detailed in Graham (2007), 
this updated relation benefits from a number of factors, including (i) the
identification of lenticular galaxies previously treated as elliptical
galaxies, 
(ii) it was constructed in the near-infrared rather than the $B$- or $V$-band
and so the 
magnitudes are less prone to biases from dust attenuation and young stellar
populations, and 
(iii) a careful S\'ersic bulge plus exponential disk decomposition has been
performed. 
Graham (2007)\footnote{Kormendy \& Gebhardt (2001) report a
$B$-band $M_{\rm bh}$-$L$ relation with a slope equal to $-0.43$. 
Graham (2007) find a consistent slope of $-0.38\pm0.06$ using the 
regression analysis from Tremaine et al.\ (2002).}, 
report that a {\it symmetrical} regression of the $M_{\rm bh}$ and $L_K$ data
yields a slope of $-0.40$.  Given that the $K$-band stellar mass-to-light
ratio is roughly constant for elliptical galaxies (Chabrier 2003; Bruzual \&
Charlot 2003), this corresponds to a one-to-one stellar-to-black hole mass
relation --- which makes sense at the high-mass end as it is the value
expected from the dry merging of galaxies and the coalescence of their SMBHs. 
The $V$-band $M_{\rm bh}$-$L$ relation used by Lauer et al.\ was much steeper than
this, having a slope of $-0.53$ (i.e.\ $M_{\rm bh} \propto L^{1.32}$), and subsequently predicted
notably more massive SMBH masses.  In addition, given that the $V$-band
stellar mass-to-light ratio increases with luminosity, an even steeper
dependence of the SMBH mass on stellar mass would be inferred, at odds with 
dry merging of massive galaxies. 
The results of applying the new $M_{\rm bh}$-$L$ and $M_{\rm
bh}$-$\sigma$ relations to predict the SMBH masses for Lauer et al.'s data are
shown in Fig.\ref{Fig7}a.
%
%

While Lauer et al.\ correctly used bulge magnitudes for the disc galaxies in
their Fig.2, these were obtained from $R^{1/4}$ bulge plus exponential disc
decompositions.  Because most bulges have a S\'ersic (1963) $R^{1/n}$ light
profile (Graham \& Driver 2005) with $n<4$ (e.g., Graham 2001; Balcells et
al.\ 2003; MacArthur et al.\ 2003), it is well known that such an 
approach overestimates the flux (e.g., Brown et al.\ 2003).  To account for
this, the bulge-to-disk flux ratios from Graham \& Driver (2007b,
their Table~2) have been applied\footnote{An extensive analysis of
  dust-corrected
bulge-to-disk systematics with disk galaxy type can be found in Graham \&
Worley (2008).}.  This entailed reducing the S0 bulge 
magnitudes by $2.5\log(0.25/0.60)$ and the Sa-Sb bulge magnitudes by
$2.5\log(0.17/0.33)$.  The results of doing this are shown in Fig.\ref{Fig7}b.  
This resolves the conflict seen in Fig.\ref{Fig7}a at the low-mass end.  While
such a correction is okay in a statistical sense, ideally individual galaxy 
corrections should be applied and this may well account for the increased
scatter about the one-to-one line in Fig.\ref{Fig7}b.

Although the BCG tend to have ($M_{\rm bh}$-$L$)-derived black hole masses
smaller than $4\times 10^9 M_{\sun}$, the Lauer et al.\ 
BCG magnitudes do tend to produce SMBH mass estimates that are roughly twice 
as large as those predicted from their velocity dispersions. 
From Liu et al.'s (2008) figure~5, one can see that the stellar envelope which
surrounds (some) BCGs becomes significant (albeit relative to an 
$R^{1/4}$ model) at $\mu_{r^{\prime}} \sim 23$ mag
arcsec$^{-2}$, while Gonzalez et al.\ (2005, their Figure~3) indicates a value
around 23.5 to 25 ${r^{\prime}}$-mag arcsec$^{-2}$.  These ranges are in 
agreement with the values seen in Seigar et al.\ 2007. 
%
This halo of stars is very likely, at least in part, due to stars that have
been tidally stripped from galaxies within the cluster environment (e.g.\
Merritt et al.\ 1985).  As such, it pertains more to the cluster
than the BCG, and should be excluded from measurements of the BCG luminosity.



To avoid the issue of the outer envelope --- which was thought to occur at
$\mu_{r^{\prime}} \sim 25$ mag arcsec$^{-2}$ --- Lauer et al.'s BCG magnitudes
were obtained from $R^{1/4}$ model fits to surface brightness profiles
brighter than $\mu_{r^{\prime}} = 23.74$ mag arcsec$^{-2}$ (Graham et al.\
1996, using $r^{\prime}-R_c = 0.24$).  The danger is that some of the
outermost portion of the light profile which was modeled may have been
elevated to a brighter level by the flux of the envelope.  As the light
profiles did not extend to large radii, the `break' in the profile, where the
envelope starts to dominate, may have been missed.  If so, such contamination
would result in the best-fitting $R^{1/4}$ model having an increased effective
radius and a brighter total flux.  Therefore, before concluding a problem
exists with the BCGs, it would be prudent to actually perform a
galaxy/envelope decomposition (e.g.\ Gonzalez et al.\ 2005; Seigar et al.\
2007) of such systems (not just those in Lauer et al.), enabling one to
quantify how the envelope flux might be biasing the magnitudes.  This is
however beyond the scope of this paper.


It is perhaps worth noting that a few galaxies are known to possess both a
SMBH and a nuclear star cluster, leading one to wonder whether a) the combined
masses should be used and b) do the `offset' barred galaxies have significant
nuclear star clusters?  As noted in GD07, the barred galaxy NGC~3384 has
$M_{\rm NC}/M_{\rm bh} \sim 2$.  Using $M_{\rm NC} + M_{\rm bh}$ rather than
only $M_{\rm bh}$ would bring this galaxy back in line with the $M_{\rm
bh}$-$\sigma$ relation.  However the barred galaxy NGC~1023 has an offset of
nearly $-0.7$ dex but $M_{\rm NC}/M_{\rm bh} \sim 0.1$, while the barred
galaxy NGC~2778 has no nuclear cluster but is offset by $-0.9$ dex.
Furthermore, the unbarred galaxy 
NGC~7457, which has no offset from the $M_{\rm bh}$-$\sigma$ relation  
has $M_{\rm NC}/M_{\rm bh} \sim 10$ (GD07).  
Nonetheless, a careful quantitative analysis of the nuclear structure of
galaxies with SMBHs would be highly desirable; it would additionally enable
one to explore the $M_{\rm bh}$ -- central surface brightness relation
proposed by GD07.

At 100 km s$^{-1}$ the barless $M_{\rm bh}$-$\sigma$ relation yields masses
(for unbarred galaxies) which are 67\% higher than the expression in Tremaine
et al.\ (2002).
Past efforts to measure the SMBH mass function and mass density using the old
$M_{\rm bh}$-$\sigma$ relation (see the roundup in Graham \& Driver 2007b,
their table~3) may therefore need tweaking.
Consequences for the $M_{\rm NC}$-$\sigma$ (and $M_{\rm NC}$-$L$) 
relation involving nuclear star clusters (e.g., Ferrarese et al.\ 2006;
Balcells et al.\ 2007) are also deferred for elsewhere.
%

\acknowledgements
The Author hereby thanks Alessandro Marconi and Leslie Hunt for kindly
providing the major-to-minor axis ratios from their GALFIT bulge models.
This research has made use of the NASA/IPAC Extragalactic Database (NED) and
HyperLeda.
This paper has benefited from two simultaneous referees, one of whom suggested
that galaxies with partially edge-on bars may have their
bulge $R_{\rm e}$ values underestimated.

\clearpage

\begin{deluxetable}{llcrc}
\tablewidth{0pt}
\tablecaption{Radii and surface brightnesses \label{Tab1}}
\tablehead{
\colhead{Galaxy} & \colhead{Type} & \colhead{$b/a$}  & \colhead{$R_{\rm e}$} & 
\colhead{$\langle \mu \rangle _{\rm e,K}$} 
}
\startdata
\multicolumn{2}{c}{`Group 1'} & & & \\
NGC 4258   & Sp,bar     & 0.51  &  0.66 &  15.26  \\
NGC 4486 (M87) & E & 0.87  &  6.0  &  16.86  \\
NGC 3115   & S0         & 0.38  &  2.9  &  16.49  \\
NGC 4649   & E     & 0.80  &  7.2  &  17.08  \\
NGC 3031 (M81) & Sp     & 0.76  &  3.0  &  16.83  \\  
NGC 4374 (M84) & E & 0.91  &  7.8  &  17.35   \\
NGC 221 (M32) & S0(?)   & 0.71  & (0.20) & (15.30) \\
NGC 5128 (Cen A) & S0   & 0.85  &  3.3  &  16.68  \\
NGC 4697   & E     & 0.58  &  6.9  &  18.18  \\
IC 1459    & E     & 0.75  &  7.1  &  16.95  \\     
NGC 5252   & S0         & 0.48  &  6.7  &  17.19  \\
NGC 2787   & S0,bar     & 0.70  &  0.27 &  14.41  \\
NGC 4594   & Sp         & 0.63  &  4.0  &  16.21  \\
NGC 3608   & E     & 0.82  &  3.9  &  17.44  \\
NGC 3245   & S0         & 0.61  &  1.0  &  15.32  \\
NGC 4291   & E     & 0.77  &  2.0  &  16.22  \\
NGC 3377   & E     & 0.51  &  3.9  &  17.91  \\  
NGC 4473   & E     & 0.57  &  2.1  &  16.41  \\
Cygnus A   & E     & 0.86  & 28.7  &  18.78  \\
NGC 4261   & E     & 0.80  &  5.8  &  16.82  \\
NGC 4564   & S0         & 0.66  & (2.4) & (17.12) \\
NGC 4742   & E     & 0.63  &  1.6  &  16.59  \\
NGC 3379   & E     & 0.87  &  2.7  &  16.54  \\
NGC 1023   & S0,bar     & 0.80  &  1.1  &  15.23  \\
NGC 5845   & E     & 0.70  &  0.42 &  13.70  \\
NGC 3384   & S0,bar     & 0.94  &  0.48 &  14.36  \\
NGC 6251   & E     & 0.82  &  10.0 &  17.06  \\
\multicolumn{2}{c}{`Group 2'} & & & \\
Milky Way  & Sp,bar     & 1.00  &  0.7  &  15.54  \\
NGC 224 (M31) & Sp      & 1.00  &  1.0  &  15.77  \\
NGC 1068   & Sp         & 0.89  &  2.9  &  15.91  \\
NGC 4459   & S0         & 0.83  &  13.7 &  19.76  \\
NGC 4596   & S0,bar     & 0.87  &  1.5  &  15.66  \\
NGC 7457   & S0         & 0.64  &  3.8  &  19.70  \\
NGC 4342   & S0         & 0.52  &  0.21 &  14.48  \\
NGC 821    & E     & 0.62  &  15.7 &  19.78  \\
NGC 2778   & S0     & 0.77  & (2.6) & (17.69) \\
NGC 7052   & E     & 0.57  &  9.1  &  17.32  \\
\enddata
\tablecomments{
Column 3: Spheroidal component's minor-to-major axis ratio, $b/a$, 
for the galaxies from Marconi \& Hunt (2003, priv.\ comm.). 
Columns 4 and 5: 
Geometric mean radii $ \left( R_{\rm e}=\sqrt{R_{\rm e,maj}^2(b/a)} \right) $ 
in kpc, and mean $K$-band effective surface brightnesses (Eq.\ref{Eq_Mu}) in mag
arcsec$^{-2}$, for the spheroidal component. 
A bracketed entry reflects that no ($K$-band) 
bulge/disk decomposition was performed.  
}
\end{deluxetable}

\clearpage


\begin{deluxetable}{lllllr}
\tablewidth{450pt}
\tablecaption{New and updated SMBH data \label{Tab3}}
\tablehead{
\colhead{Galaxy} & \colhead{Type}  &  \colhead{Dist.}  &  
\colhead{$\sigma$}  &  \colhead{$\log R_{\rm e}$[kpc]}  & 
 \colhead{$M_{\rm bh}$}  \\
\colhead{} &  \colhead{} &  \colhead{(Mpc)}  &
  \colhead{km s$^{-1}$}  & \colhead{}  &  \colhead{($10^8 M_{\sun}$)} \\
\colhead{1} &  \colhead{2} &  \colhead{3}  &
  \colhead{4}  & \colhead{5}  &  \colhead{6}  
}
\startdata
\multicolumn{6}{c}{Updated Marconi \& Hunt (2003) data}  \\
Cygnus A  & E    &  240          &  270    &  1.49        &  $26.0^{+7.0}_{-7.0}$ [2] \\
Cen A     & S0   &  4.2          &  150    &  0.56        &  $1.1^{+0.1}_{-0.1}$ [10] \\ 
NGC 221   & S0   &  0.8          & 75      & $-0.98$ [6] &  $0.025^{+0.005}_{-0.005}$  \\
NGC 821   & E    & 24.1          & 200 [5] & 1.30         &  $0.85^{+0.35}_{-0.35}$ [11] \\
NGC 2778  & S0   & 22.9          & 175     & $-0.60$ [2]  &  $0.14^{+0.08}_{-0.09}$ \\
NGC 3379  & E    &  10.6         &  206    &  0.46        &  $1.4^{+2.6}_{-1.0}$ [12] \\
NGC 4342  & S0 & 17.0 [1]        & 251 [5] & $-0.36$      &  $3.3^{+1.9}_{-1.1}$  \\
NGC 4374  & E    &  18.4         &  296    &  0.91        &  $4.64^{+3.46}_{-1.83}$ [13] \\
NGC 4564  & S0   &  15.0         & 162     & $-0.50$ [2]  &  $0.56^{+0.03}_{-0.08}$  \\
NGC 5252  & S0 & 94.4 [2]        & 190     &  0.98        &  $9.7^{+14.9}_{-4.6}$  \\
NGC 6251  & E    &  101 [2]      & 290     &  1.02        &  $5.8^{+1.8}_{-2.0}$  \\
NGC 7052  & E    & 60.0 [2]      & 266     & 1.00         &  $3.4^{+2.4}_{1.3}$ \\
\multicolumn{6}{c}{New galaxies with SMBH measurements}  \\ 
NGC 1300  & SBbc & 19.3$h_{73}$  & 229 [5] & $-0.28$ [7] & $0.68^{+0.65}_{-0.33}$ [14] \\
NGC 1399  & E    & 20.0 [3]      & 317     &  1.09 [4]   & $12^{+5}_{-6}$ [15]     \\
NGC 2748  & Sbc  & 23.8$h_{73}$  & ...     &  ...        & $0.45^{+0.36}_{-0.37}$ [14] \\
NGC 3227  & SB   & 17.0$h_{73}$  & 160     & $-0.57$     & $0.15^{+0.05}_{-0.08}$ [16] \\
NGC 3998  & S0   & 14.1 [3]      & 305     & $-0.16$     & $2.2^{+2.0}_{-1.7}$ [17] \\
NGC 4151  & SBab & 13.9          & 156 [5] & $-0.23$ [8] & $0.45^{+0.05}_{-0.05}$ [18] \\
NGC 4435  & SB0  & 16.0          & 157     & $-0.07$     &  $<0.075$ [19] \\
NGC 4486a & E    & 17.0 [1]      & 110     & $-0.39$ [9] & $0.13^{+0.08}_{-0.08}$ [20] \\
NGC 7582  & SBab & 22.4          & 156     & ...         & $0.55^{+0.26}_{-0.19}$ [21] \\
%
%
%
%
\enddata
\tablecomments{
Unless otherwise specified, the distances, velocity dispersions, and 
effective radii of each galaxy have come from the reference which provides the
SMBH mass.  Both $M_{\rm bh}$ and $R_{\rm e}$ have been adjusted to the
distance given in column~2. 
References: 
 1 = Jerjen et al.\ (2004); 
 2 = Graham \& Driver (2007a); 
 3 = Tonry et al.\ (2001); 
 4 = D'Onofrio et al.\ (1994); 
 5 = HyperLeda (http://leda.univ-lyon1.fr/); 
 6 = Graham (2002);
 7 = Aguerri et al.\ (2001);
 8 = Virani et al.\ (2000);
 9 = Kormendy et al.\ (2005);
 10 = Marconi et al.\ (2006); 
 11 = Richstone et al.\ (2007); 
 12 = Shapiro et al.\ (2006); 
 13 = Maciejewski \& Binney (2001); 
 14 = Atkinson et al.\ (2005); 
 15 = Houghton et al.\ (2006); 
 16 = Davies et al.\ (2006); 
 17 = De Francesco et al.\ (2006); 
 18 = Onken et al.\ (2007);
 19 = Coccato et al.\ (2006); 
 20 = Nowak et al.\ (2007);
 21 = Wold et al.\ (2006).
}
\end{deluxetable}

\begin{deluxetable}{lll}
\tablewidth{450pt}
\tablecaption{Galaxies with somewhat uncertain $M_{\rm bh}$ \label{Tab4}}
\tablehead{
\colhead{Galaxy} & \colhead{Reference}  &  \colhead{Note}  
}
\startdata
Abell 1836 & Dalla Bont\`a et al.\ 2007  &  no refereed publication \\
Abell 3565 & Dalla Bont\`a et al.\ 2007  &  no refereed publication \\
Circinus &  Greenhill et al.\ 2003  &  poorly known disk inclination \\
IC 1459  &  Cappellari et al.\ 2002  &  gas/stellar dynamics differ \\
NGC 1068 &  Hur{\'e} 2002; Lodato \& Bertin 2003  &  disk model uncertain \\ 
NGC 4041 &  Marconi et al.\ 2003  &  disk might be dynamically decoupled \\
NGC 4303 &  Pastorini et al.\ 2007  & poorly known disk inclination \\
NGC 4350 &  Pignatelli et al.\ 2001  &  possible SMBH, high $M_{\rm bh}/M_{\rm bulge}$ \\ 
NGC 4459 &  Sarzi et al.\ 2001  &  poorly known disk inclination \\
NGC 4486B & Kormendy et al.\ 1997  &  possible SMBH, $M_{\rm bh}/M_{\rm bulge} = 0.09$ \\ 
NGC 4594 &  Kormendy 1988 & no 3-integral model \\
NGC 4596 &  Sarzi et al.\ 2001  &  poorly known disk inclination \\
NGC 4945 &  Greenhill et al.\ 1997  &  no 2D velocity field \\
NGC 5055 &  Blais-Ouellette et al.\ 2004  &  possibly no black hole \\
NGC 7332 &  H\"aring \& Rix 2004  &  no refereed publication \\ 
\enddata
\end{deluxetable}

\clearpage

\begin{deluxetable}{lccc}
\tablewidth{490pt}
\tablecaption{SMBH mass -- spheroid relations \label{Tab5}}
\tablehead{
\colhead{Sample} & \colhead{Relation}  &  \colhead{$\Delta$} &  \colhead{$\Delta^{\prime}$} \\
\colhead{ } & \colhead{}  &  \colhead{dex} &  \colhead{dex} 
}
\startdata
              & \colhead{$M_{\rm bh}$-$\sigma$}  &   & \\
40 galaxies   & $8.13\pm0.06 +(3.92\pm0.27)\log(\sigma/200)$ &  0.38  & 0.35 \\
29 non-barred & $8.26\pm0.06 +(3.67\pm0.19)\log(\sigma/200)$ &  0.30  & 0.25 \\
19 elliptical & $8.25\pm0.05 +(3.68\pm0.25)\log(\sigma/200)$ &  0.24  & 0.18 \\
\hline
              & \colhead{$M_{\rm bh}$-$\sigma$-$R_{\rm e}$}  &  &  \\
40 galaxies   &  $8.19\pm0.05 +(3.23\pm0.28)\log(\sigma/200) +
(0.43\pm0.11)[\log(R_{\rm e}/3)]$ &  0.30 & 0.28 \\
29 non-barred &  $8.26\pm0.05 +(3.29\pm0.26)\log(\sigma/200) +
(0.29\pm0.11)[\log(R_{\rm e}/3)]$ &  0.25 & 0.24 \\
19 elliptical &  $8.23\pm0.04 +(3.32\pm0.36)\log(\sigma/200) +
(0.21\pm0.15)[\log(R_{\rm e}/3)]$ &  0.22 & 0.18
\enddata
\tablecomments{The total scatter $\Delta$ is given 
rather than the (smaller) internal/intrinsic scatter, as the latter quantity 
depends on the measurements errors that one assigns. The final column shows
the total scatter $\Delta^{\prime}$ after removing just two data points (Cygnus~A and
NGC~3998).} 
\end{deluxetable}

\clearpage

\clearpage

\clearpage

\end{document}